\documentclass[9pt,sigconf,letterpaper]{acmart}

\usepackage{natbib} 
\usepackage{algorithmic}
\usepackage{graphicx}
\usepackage{textcomp}
\usepackage{tumcolors}
\usepackage{booktabs}
\usepackage{todonotes}
\usepackage{url}
\usepackage{enumitem}
\usepackage{siunitx}
\usepackage{multirow}
\usepackage{caption}
\usepackage{subcaption}
\usepackage{soul}

\newcommand{\first}{\emph{(i)}~}
\newcommand{\second}{\emph{(ii)}~}
\newcommand{\third}{\emph{(iii)}~}

\DeclareSIUnit{\bit}{b}

\setitemize{leftmargin=*}
\copyrightyear{2023} 
\acmYear{2023} 
\setcopyright{acmlicensed}
\acmConference[ANRW '23]{Applied Networking Research Workshop}{July 24, 2023}{San Francisco, CA, USA}
\acmBooktitle{Applied Networking Research Workshop (ANRW '23), July 24, 2023, San Francisco, CA, USA}
\acmPrice{15.00}
\acmDOI{10.1145/3606464.3606474}
\acmISBN{979-8-4007-0274-7/23/07}

\begin{document}

\title{Evaluating the Benefits: Quantifying the Effects of TCP Options, QUIC, and CDNs on Throughput}

\author{Simon Bauer}
\affiliation{%
  \institution{Technical University of Munich}
    }
\email{bauer@net.in.tum.de}

\author{Patrick Sattler}
\affiliation{%
  \institution{Technical University of Munich}
    }
\email{sattler@net.in.tum.de}

\author{Johannes Zirngibl}
\affiliation{%
  \institution{Technical University of Munich}
    }
\email{zirngibl@net.in.tum.de}

\author{Christoph Schwarzenberg}
\affiliation{%
  \institution{Technical University of Munich}
    }
\email{schwarzenberg@net.in.tum.de}

\author{Georg Carle}
\affiliation{%
  \institution{Technical University of Munich}
    }
\email{carle@net.in.tum.de}

\begin{abstract}

To keep up with increasing demands on quality of experience, assessing and understanding the performance of network connections is crucial for web service providers.
While different measures, like TCP options, alternative transport layer protocols like QUIC, or the hosting of services in CDNs, are expected to improve connection performance, no studies are quantifying such impacts on connections on the Internet.

This paper introduces an active Internet measurement approach to assess the impacts of mentioned measures on connection performance. 
We conduct downloads from public web servers considering different vantage points, extract performance indicators like throughput, RTT, and retransmission rate, and survey speed-ups due to TCP option usage. 
Further, we compare the performance of QUIC-based downloads to TCP-based downloads considering different option configurations. 

Next to significant throughput improvements due to TCP option usage, in particular TCP window scaling, and QUIC, our study shows significantly increased performance for connections to domains hosted by different giant CDNs.

\end{abstract}


\maketitle

\section{Introduction}

Due to its impact on user satisfaction, understanding the performance of connections and the impact of potential performance improvements is crucial for service and infrastructure providers. 
The same applies from a research perspective to assess the effectiveness of arising or widely deployed measures to improve connection performance. 

The transmission control protocol (TCP), responsible for the majority of Internet traffic\,\cite{bauer2021evolution}, was extended for several options proposed to improve the performance shortcomings of its original design. 
Further, QUIC and HTTP3 represent an arising alternative to the commonly used TCP/HTTPS stack\,\cite{mucke2022waiting}.
This development also motivates a closer look at performance differences between QUIC and TCP connections targeting the same resources in productive deployments, i.e., on the Internet. 
Next to protocol usage and their configuration, the hosting of service infrastructure is crucial to ensure availability and performance.
This led to the trend towards a more centralized Internet, i.e., more services being hosted in large-scale content delivery networks (CDNs)\,\cite{gerber2011traffic,poese2010improving,GigisOffnets2021,zembruzki2022centralization}.

However, while there are publications surveying deployments and usage of different TCP options\,\cite{kuhlewind2013state,murray2017analysis, lim2022fresh}, comparing the performance of TCP and QUIC connections in controlled test environments\,\cite{nepomuceno2018quic,biswal2016does}, or focusing on optimizations of both protocol \\ stacks~\,\cite{wolsing2019performance}, there are no insights on the impact of TCP option usage, QUIC usage, or CDN hosting on the performance of Internet connections.

This paper assesses the impacts of the named measures on connection performance by conducting active measurements with public Internet servers. 
Thereby, we exploit the capability of active Internet measurements to determine client configurations and target selection.

Our contributions in this work are:

\first{}
We introduce an active measurement approach for public Internet web servers covering crawling of suitable measurement targets, conducting downloads with different client configurations, and analyzing the performance of connections by extracting different performance indicators. 

\second{}
We apply the introduced approach to measurements from different vantage points on a set of publicly available web servers chosen from Internet top lists and discuss corresponding measurement results in this paper. 

\third{}
We publish the implemented measurement pipeline. 
\section{Background}

\paragraph{Performance-related TCP Options}
In this paper, we consider three options addressing the performance of TCP connections: window scaling (WS), selective acknowledgments (SACK), and explicit congestion notifications (ECN). 
TCP window scaling\,\cite{rfc1323} is proposed to exceed limitations of bytes in flight implied by the length of the \texttt{window} field in the TCP header from \SI{65}{\kilo\byte} up to \SI{1}{\giga\byte}.
TCP WS enables the exchange of a factor during connection establishment to shift all following values of the \texttt{window} field.

Selected acknowledgments\,\cite{rfc2018} are purposed to avoid unnecessary retransmissions by specifying lost packets.
This is achieved by exchanging ranges of sequence numbers of successfully received packets. %
Today, WS and SACK are enabled on all major operating systems like Linux, MacOS, or Windows\,\cite{murray2017analysis}. 

Explicit congestion notifications (ECN)\,\cite{rfc3168} are a measure to avoid overload on the network and, accordingly, lost packets by enabling routers to signal congestion actively. 
As routers do not access Layer 4 headers, ECN relies on two bits each in a packet's IP and TCP header. 
Permutations of such flags are then used to signal ECN support, detected congestion, and reaction to observed congestion.
In addition to the endpoints, all routers of a network path have to support ECN.
Typical operating systems support the usage of ECN for at least incoming connections\,\cite{lim2022fresh}. 

Note that we do not consider TCP Fast Open (TFO)\,\cite{rfc7413} for this study. 
TFO is particularly effective if a client requests several sources from a server, resulting in several TCP handshakes implying overhead. 
This use case does not match our approach to explicitly download single files from a target, as described in Section\,\ref{sec:approach}. 

\paragraph{QUIC}
QUIC is a transport layer protocol specified in 2021\,\cite{rfc9000,rfc9001,rfc9002}, providing properties like reliable data transmission, connection migration, and encryption while relying only on UDP packet sequences. 
QUIC implements selective acknowledgments by acknowledging ranges of packet numbers indicating lost packets.
In contrast to TCP, which limits selective acknowledgments to a maximum of three ranges of sequence numbers, QUIC supports up to 256 ranges.
Further, QUIC supports ECN usage. 
Upper bounds for receiver window sizes differ between QUIC implementations.
\section{Related Work}

TCP options and their deployment are frequently addressed topics. 
Studies of TCP option deployments conducted in the early 2000s observed the evolution of option deployment, starting from only small adoption of servers to TCP options\,\cite{allman2000web, PentikousisOptionDepl, medina2004measuring}.
A study conducted in 2013 by K{\"u}hlewind et al.\,\cite{kuhlewind2013state} reports widespread deployment of WS and SACK and observes a slower spreading of ECN usage.
Such observation is confirmed by Murray et al.\,\cite{murray2017analysis} in 2017, who only observed small usage of ECN in captured Internet traffic from a university network.
More recent studies observe that ECN is used by the majority of domains listed in the Alexa Top 1 M list\,\cite{chen2019analysis}, respectively, in passively captured university network traffic\,\cite{lim2022fresh}. 
The interference of middleboxes on TCP options is surveyed by Honda et al.\,\cite{honda2011still}.
Edeline and Donnet\,\cite{edeline2020evaluating} survey the impact of TCP option usage in controlled test environments showing the beneficial effects of TCP options.

According to W3Techs~\cite{w3techquic} QUIC accounted for 8\% of the total global Internet traffic in 2022.
Shreedhar et al.\,\cite{shreedhar2021evaluating} compare QUIC to the TCP/TLS stack and observe significantly smaller connection duration for web workloads on the Internet.
However, TCP option usage is not considered by the study.
Further publications show that QUIC outperforms TCP in different controlled test environments\,\cite{biswal2016does, yu2017quic, nepomuceno2018quic}. 

Additional studies survey quality of experience (QoE) metrics of different web applications based on passive data sets \,\cite{mangla2018emimic, lopez2019effective, madanapalli2019inferring}. 
In contrast, our work analyzes transport layer performance based on active measurements.
The reproduction of realistic web applications and web pages for performance measurements was studied by Jun et al.\,\cite{Jun2021webtune} and Zilberman et al.~\cite{zilbermannrg} 

Considering the above state of the art, there are only limited insights into the impact of TCP option usage on the performance of Internet connections, while TCP options are commonly deployed.
The same applies to the implications of QUIC usage and the impacts of CDN hosting on connection performance. %

\section{Approach}%
\label{sec:approach}

For our study, we download files provided by public web servers taken from Internet top lists with varying TCP options and QUIC. 
This section describes the different steps of our active measurement approach, like determining and selecting suitable measurement targets and conducting downloads with controlled client configurations.
Considered performance indicators and other extracted metrics are introduced in Section \ref{sec:impl}.

\subsection{Determining Measurement Targets}

Conducting active measurements with public and uncontrolled targets on the Internet requires crawling domains for suitable files to be downloaded. 
We refer to a suitable file if it satisfies a specific minimum file size, purposed to provide comparable results between different domains.
For our study, we choose a minimum file size of \SI{1}{\mega\byte}. 
While this is a relatively large file size considering the distribution of flow sizes observed in passive data sets\,\cite{bauer2021evolution}, the same study emphasizes the relevance of connections with such size regarding the totally observed bytes. 

Based on suitable target domains and corresponding files, we compose a target set considering six different groups regarding CDN hosting.
Firstly we consider domains hosted by different giant CDNs, i.e., Akamai, Amazon, Cloudflare, Google, and Microsoft (200 domains for each).
Secondly, a sixth target group consisting of targets not hosted by the listed organizations (1000 domains).
We map domains to the selected CDN providers based on the IP addresses observed for a domain, their mapping to the announcing autonomous system (AS) based on BGP dumps from a Route Views~\cite{routeviews} collector, and a mapping of ASes to their respective organizations based on the work from Arturi et al.~\cite{arturi2023as2orgplus}.
We ensure that all selected domains support the three considered TCP options. 

\subsection{Conducting Downloads}

Next, we initiate downloads of the crawled files for each domain in the composed target set. 
We consider different permutations of TCP option usage for downloads and conduct one download for each configuration for a domain sequentially.
Before each download, we freshly resolve the target domain to ensure adaption to DNS-based load balancing.
Further, we conduct a warm-up run (TCP, no option enabled) to avoid bias between downloads of the same measurement run due to edge caching by CDNs. 
After conducting TCP downloads, we run downloads of the same file with different QUIC clients as elaborated in Section\,\ref{sec:impl}.
Afterward, we continue with downloads from the following domain in the target list.

We examine a specific collection of TCP option configurations, which include: (i) a baseline configuration (BL) that does not utilize any options, (ii) configurations supporting only one of the considered options (ECN, SACK, WS), and (iii) a configuration that enables all options (ALL).
We use the maximum window scaling factor of 14 for all listed configurations supporting the option.
To allow a fair comparison between TCP and QUIC, we only conduct TCP downloads via HTTPS, implying that TCP and QUIC connections provide encrypted data transmission. 

By design, this measurement approach is limited to control configuration and conditions at the client. 
This implies that the server and its characteristics, like operating systems, server implementations, or used congestion control algorithms, are not known. 
The same applies to load conditions on the Internet paths and at the target server.
We conduct downloads with different configurations back to back for one domain, referred to as one measurement run.
This procedure ensures that conditions in the network and on the server side are as similar as possible for all downloads of one run, e.g., regarding daytime patterns of service usage and corresponding load. 
Further, measured performance indicators are compared within one measurement run, for instance, to calculate speed-ups by a configuration.

\subsection{Vantage Points}

As connection performance also depends on the location of the vantage point (VP), considering the distance to target servers and last-mile network conditions, we use three different vantage points for our measurements: 
First, a physical server located in a campus data center in Munich (MUC) and, second, two virtual machines hosted by the cloud provider DigitalOcean in data centers located in San Francisco (SFO) and Singapore (SGP).
The physical server hosted in Munich is connected with a \SI{1}{\giga\bit\per\second} up- and downlink to the German science network (DFN) that connects to the Internet via a major Tier 1 provider.
The measurement host is equipped with an Intel Xeon E5-2630 CPU providing six physical cores at a clock frequency of \SI{2.6}{\giga\hertz}, \SI{32}{\giga\byte} memory, and a Broadcom NetXtreme BCM5719 Gigabit NIC.
The virtual machines hosted in SFO and SGP are equipped with two virtual CPU cores and 4 GB memory. 

\subsection{Ethical Considerations}
Active measurements on public infrastructure like the Internet require responsible measurement practices.
We followed a set of ethical measures, i.e., informed consent~\cite{menloreport} and community best practices~\cite{PA16} during all our scans.
Our measurement hosts' IP addresses can be identified via reverse DNS or WHOIS information, while the measurement host operates an explanatory website.
We maintain an abuse contact email and react quickly to all requests, including the option to exclude a domain or IP range from further measurements.
We use a custom HTTP user agent to be identifiable as a research group and follow crawling instructions in the robots.txt according to the Robots Exclusion protocol~\cite{rfc9309}. 
\section{Implementation}
\label{sec:impl}

This section describes the implementation of the different measurement pipeline components. 
The implemented pipeline is publicly available\,\cite{code:repo}. 

\subsection{Crawling and Conducting Downloads} 
Based on a set of domains, crawling aims to identify web servers providing suitable files, as described in Section\,\ref{sec:approach}. 
In order to determine files for downloads, the crawler recursively follows links found on a crawled website if links explicitly point to the same domain and can be reached by the same IP address as the initially crawled website. 
The crawling component sends HTTP HEAD requests to extract the optional HTTP Content-Length field\,\cite{fielding2014rfc} with the Python crawling library \textit{Scrapy} to determine the size of a crawled file.
First tests showed that many targets do not provide Content-Length information.
Accordingly, we implemented a fallback by starting downloads of files and stopping them if the minimum file size was successfully downloaded.

Crawling targets results in a list of domains and corresponding files suitable for our downloads.
The download component iterates over the list of determined files and conducts a download with each specified TCP configuration, respectively QUIC. 
Before downloads are established, the downloader resolves the target domain's IP address to ensure an up-to-date resolution. 
Afterward, an HTTPS GET request is sent to the freshly resolved target IP with the python \textit{HTTP Requests} library. 
We use the \textit{ForcedIPHTTPS adapter}\,\cite{forcediphttpsadapter:website} to enable the use of specific IP addresses while establishing a TLS/SSL connection. 
Download traffic gets captured with \textit{tcpdump}.

The downloader relies on the corresponding settings of the Linux kernel to configure TCP option usage. 
In particular, the downloader sets flags indicating ECN, SACK, and WS usage and sets the kernel's TCP receive memory size to enforce the use of a specific window scaling factor. 
To conduct QUIC downloads, we rely on the QUIC implementations \textit{aioquic}\,\cite{impl_aioquic} and \textit{quiche}\,\cite{impl_quiche-cloudflare}.
Considering different QUIC implementations is motivated by a recently conducted study comparing the performance of QUIC implementations in controlled test environments\,\cite{Jaeger2023QUIC}. 
According measurement results show that \textit{quiche} outperforms \textit{aioquic} in high bandwidth scenarios.
This observation motivates to survey whether such performance differences are also represented in downloads conducted on the Internet. 
Integrating additional QUIC implementations is a considered extension of our measurement pipeline.

\subsection{Traffic Analysis}

For our study, we primarily examine the throughput of connections.  
We calculate average throughput, in the following referred to as mean throughput, as the fraction of transmitted data and the duration of a connection.
The amount of transferred data is determined by the sum of packet sizes of a connection, as specified by the \textit{total length} field in IP packet headers.

In addition, the traffic analysis component also extracts performance indicators like mean round trip time (RTT), total retransmission rate (RR), or goodput to further survey the performance characteristics of analyzed connections.  
We calculate the RTT based on the TCP timestamp option, which enables matching tuples of corresponding packets based on \textit{TSval} and \textit{TSecr} values.
The difference of corresponding packet timestamps then determines an RTT sample. 
We determine packets as retransmissions according to the Wireshark documentation considering retransmissions, fast retransmissions, and spurious retransmissions\,\cite{wireshark_doc}.
The retransmission rate is then calculated as the fraction of data transported by retransmitted packets and the total number of observed data. 
Goodput is calculated as the fraction of the sum of packet sizes without considering retransmissions and the duration of a connection.
Further, we extract different IP and TCP header fields to survey the effective use of TCP options like ECN echo and CWR flags or SACK blocks from the TCP header.
For QUIC connections, we only consider mean throughput as a performance indicator, calculated in the same manner as for TCP. 

\begin{table}
    \center
    \caption{Number of domains resulting in successful downloads.}
    \begin{tabular}{lrrrrrrr}
        \toprule
        \rotatebox{75}{Run} &  \rotatebox{75}{Total} &  \rotatebox{75}{Akamai} &  \rotatebox{75}{Amazon} & \rotatebox{75}{Cloudflare} &  \rotatebox{75}{Google} &  \rotatebox{75}{Microsoft} &  \rotatebox{75}{Others} \\
        \midrule
        \multicolumn{8}{c}{\bfseries TCP} \\
        \midrule
        MUC &  1679 & 167 & 150 & 170 &    147 &   172 &    873 \\
        SFO &  1678 & 165 & 147 & 168 &    152 &   173 &    873 \\
        SGP &  1640 & 162 & 147 & 163 &    143 &   166 &    859 \\
        \midrule
        \multicolumn{8}{c}{\bfseries QUIC} \\
        \midrule
         MUC\_Q &   511 &   3 &  15 & 289 &      2 &     0 &    202 \\
         SFO\_Q &   506 &   3 &  14 & 285 &      2 &     0 &    202 \\
         SGP\_Q &   495 &   3 &  13 & 276 &      2 &     0 &    201 \\
        \bottomrule
    \end{tabular}
    \label{tab:targets:stats}
\end{table}

\section{Internet Measurements}
\label{sec:results}

This section evaluates performance indicators extracted during conducted downloads. 
In addition to connection performance, this section surveys the option deployment across domains in the Alexa Top 1M list.  

\subsection{Option Deployment}

To provide a recent view on option deployment on the Internet, we conduct active measurements to all domains included in the Alexa Top 1M list (the selected list contained 1M domains). 
Measurements are conducted by establishing TCP handshakes to the index page of each domain. 
After establishing the handshake, we immediately terminate TCP connections, as the handshake is sufficient to extract supported options.
\SI{5.3}{\percent} of domains do not support a single option while \SI{81.0}{\percent} support all three considered options.
ECN is supported by \SI{85.8}{\percent}, SACK by \SI{91.4}{\percent} and WS by \SI{91.1}{\percent} of the domains.

\begin{figure} 
\centering
\includegraphics[width=0.5\textwidth]{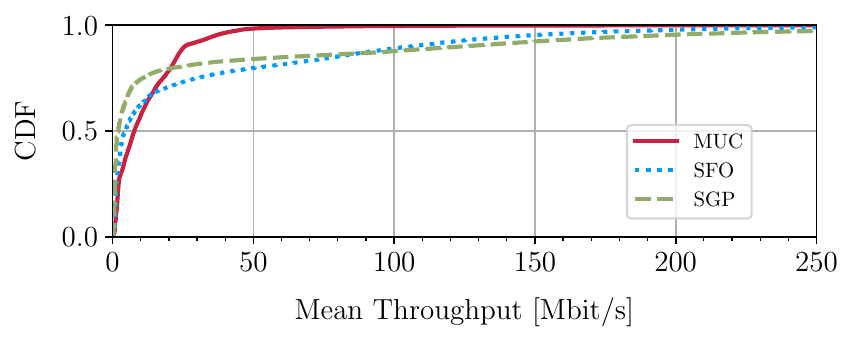} %
\includegraphics[width=0.5\textwidth]{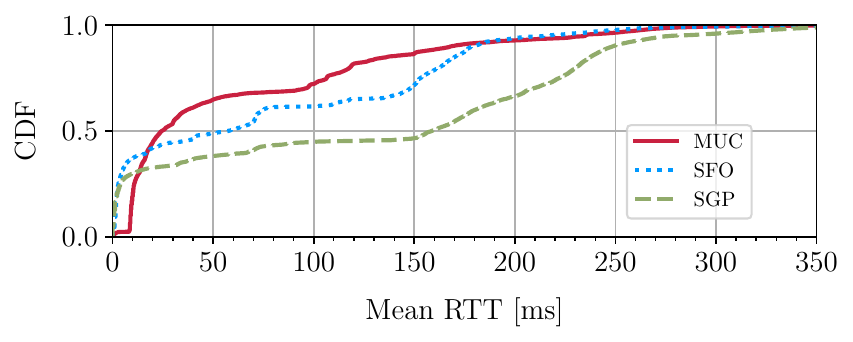}%
\caption{Measured KPIs per VP.}
\label{fig:vantage_point_kpis}
\end{figure}

\begin{figure*}
\centering
\begin{subfigure}{0.33\textwidth}
 \centering\includegraphics[width=\textwidth]{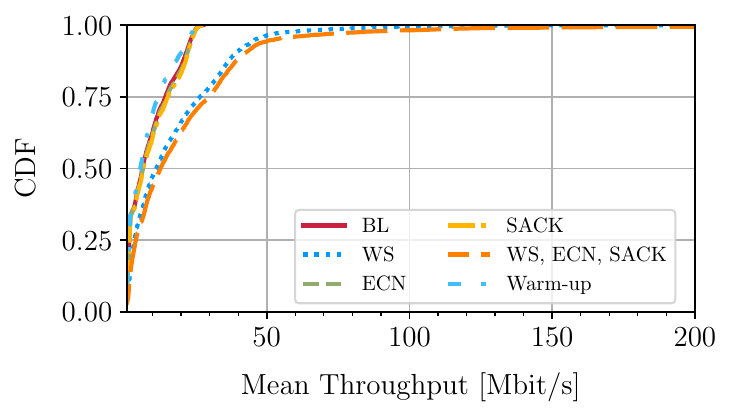}
    \caption{Vantage point in MUC.}
    \label{fig:pl27_tp_x_opts}
\end{subfigure}%
\begin{subfigure}{0.33\textwidth}
 \centering\includegraphics[width=\textwidth]{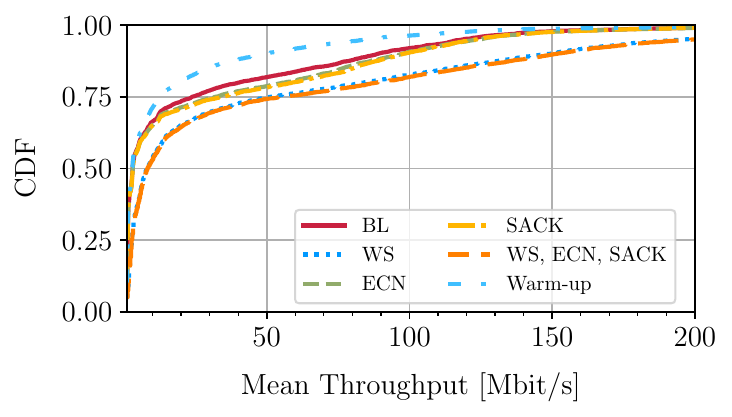}
    \caption{Vantage point in SFO.}
    \label{fig:do_tp_x_opts}
\end{subfigure}%
\begin{subfigure}{0.33\textwidth}
 \centering\includegraphics[width=\textwidth]{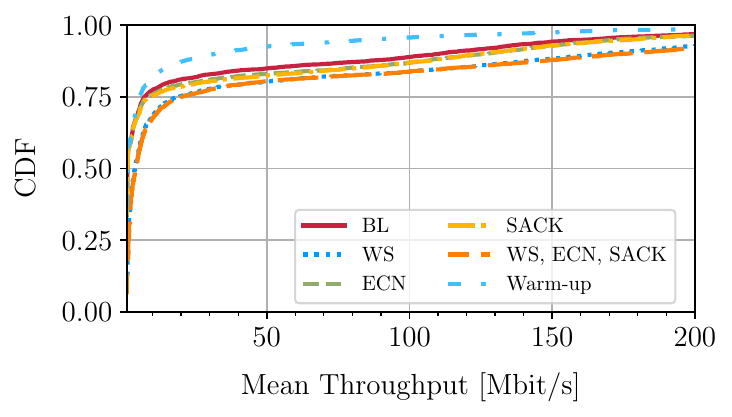}
    \caption{Vantage point in SGP.}
    \label{fig:dosg_tp_x_opts}
\end{subfigure}%
\caption{Distribution of mean throughput measured with different TCP option configurations.}
\label{fig:tcp_option_impact}
\end{figure*}

\begin{table*}[]
\center
\caption{Shares of downloads of a configuration (Config.) resulting in a certain speed-up compared to another configuration (vs.) aggregated for all VPs.}
\begin{tabular}{llccccccccccc}
\toprule
\multicolumn{13}{c}{\bfseries TCP options} \\  \midrule

Config & vs. & + & - & 0.7 - 0.8 & 0.8 - 0.9 & 0.9 - 1.0 & 1.0 - 1.1 & 1.1 - 1.2 & 1.2 - 1.3 & 1.3 - 1.5 & 1.5 - 2 & >2 \\ \midrule

Warm-up & BL & 35.4\%  & 64.6\% & 3.8\% & 8.3\% & 37.1\% & 25.9\% & 3.8\% & 1.5\% & 1.6\% & 1.2\% & 1.5\% \\
ECN    & BL & 53.3\%  & 46.7\% & 2.1\% & 5.3\% & 34.0\% & 35.0\% & 6.1\% & 2.3\% & 2.6\% & 2.2\% & 5.1\% \\
SACK   & BL & 54.2\%  & 45.8\% & 2.1\% & 5.3\% & 33.2\% & 34.7\% & 6.3\% & 2.8\% & 2.6\% & 2.4\% & 5.5\% \\
WS     & BL & 90.3\%  & 9.7\% & 0.9\% & 1.6\% & 3.5\% & 5.7\% & 6.8\% & 6.1\% & 10.8\% & 22.9\% & 38.0\% \\
All    & BL & 91.4\%  & 8.6\% & 1.0\% & 1.3\% & 3.2\% & 5.6\% & 6.8\% & 5.7\% & 10.2\% & 22.8\% & 40.2\% \\

\toprule
\multicolumn{13}{c}{\bfseries QUIC and TCP} \\  \midrule
Config. & vs.  & + & - & 0.7 - 0.8 & 0.8 - 0.9 & 0.9 - 1.0 & 1.0 - 1.1 & 1.1 - 1.2 & 1.2 - 1.3 & 1.3 - 1.5 & 1.5 - 2 & >2 \\  \midrule
quiche  & aioquic & 70.0\%  & 30.0\% & 4.0\% & 2.8\% & 2.9\% & 4.0\% & 4.6\% & 5.9\% & 7.7\% & 8.9\% & 38.9\% \\
aioquic & TCP-BL & 59.7\%  & 40.3\% & 2.4\% & 2.1\% & 2.5\% & 3.4\% & 2.4\% & 2.2\% & 6.4\% & 8.5\% & 36.7\% \\
aioquic & TCP-ALL & 44.5\%  & 55.5\% & 5.1\% & 7.1\% & 4.7\% & 3.3\% & 1.6\% & 1.1\% & 1.6\% & 4.6\% & 32.4\% \\
quiche & TCP-BL & 82.9\%  & 17.1\% & 2.1\% & 2.8\% & 3.9\% & 5.2\% & 3.9\% & 2.9\% & 4.7\% & 14.6\% & 51.5\% \\ 
quiche & TCP-ALL & 71.9\%  & 28.1\% & 4.3\% & 4.3\% & 5.5\% & 9.1\% & 7.5\% & 4.3\% & 4.3\% & 5.9\% & 40.7\%\ \\
\bottomrule

\end{tabular}
\label{tab:tcp_option_speedups}
\end{table*}

\subsection{Targets for Performance Measurements}

To compose our target set, we crawl the top 100K entries of the  Alexa Top 1M list. 
Crawling results in over 22K measurement targets providing a file of at least \SI{1}{\mega\byte}.
We select 2000 domains according to the organization maintaining the AS number of a domain, as described in Section\,\ref{sec:approach}, referred to as the TCP target set. 
We observe that not all downloads from successfully crawled domains succeed. 
Such observation can be explained by crawled files being no longer available. 
Further, we observed that a significant share of domains hosted by Cloudflare did not result in successful downloads for the vantage points in SFO and SGP during previous measurements, while downloads succeeded from the vantage point in MUC.
This observation indicates that Cloudflare blocked some of our download attempts, e.g., through human verification for selected IP ranges\,\cite{CFprotect}. 
However, we do not observe such impact during the final measurements. 

We find only negligible shares of domains resulting in successful QUIC downloads in the TCP target set. 
Therefore, we compose a second target set, referred to as the QUIC target set. 
As the Alexa Top 1M list was retired in February 2023, the QUIC target set comprises domains taken from the top 100K entries of Google's CrUX dataset\,\cite{about_crux}. 
We determine domains supporting QUIC with the \textit{QScanner} introduced by Zirngibl et al.\,\cite{zirngibl2021over9000}. 
Based on the list of domains supporting QUIC, we choose targets providing a suitable file and supporting all considered TCP options.
Finally, we merge QUIC-supporting targets from the Alexa-based TCP target set with the domains taken from the CrUX dataset.
This procedure results in 558 suitable measurement targets.  
In the future, scanning for QUIC targets might be replaced by analyzing HTTPS DNS resource records which i.a., provide information regarding a domain supporting QUIC.
However, Zirngibl et al. have shown that the record is currently used mainly by Cloudflare\,\cite{zirngibl2023svcb}. 

We run three measurement iterations for both target sets, while one iteration consists of one measurement run per domain. 
Note that measurements based on the QUIC target set only consider downloads with \textit{aioquic}, \textit{quiche}, TCP-BL, TCP-ALL, and a warm-up run.
Table\,\ref{tab:targets:stats} lists the number of domains resulting in successful downloads during the first measurement iteration, numbers of successful downloads for following iterations vary slightly within a deviation smaller than \SI{3}{\percent}.

\subsection{Comparison of Performance Indicators per Vantage Point} 

Conducting measurements from different vantage points indicates varying impacts on observed performance, for instance, due to different traversed Internet paths and distances between vantage points and measurement targets.
Therefore, we survey performance indicators independent of the client configuration for the used vantage points. 
Figure\,\ref{fig:vantage_point_kpis} shows the cumulative distribution function (CDF) of mean throughput and mean RTT for the three vantage points for all considered option configurations except warm-up runs.
The VP in MUC results in a larger mean throughput up to about the 75th percentile of samples. 
Afterward, the VPs in SFO and SGP show larger shares of downloads with significantly increased mean throughput compared to the VP in MUC. 
This observation correlates to the distribution of mean RTTs. 
In particular, we find that the VPs in SFO and SGP result in a significant share of mean RTTs smaller than \SI{5}{\milli\second}, which is not observed for the VP in MUC. 
However, the VP in MUC shows smaller mean RTTs for most of the remaining samples. 
We find that measurements conducted from SFO and SGP with mean RTTs smaller than \SI{5}{\milli\second} are associated with measurement targets hosted by Akamai, Cloudflare, and partly Amazon. %

\begin{figure*}
\centering
\begin{subfigure}{0.33\linewidth}
 \centering\includegraphics[width=\linewidth]{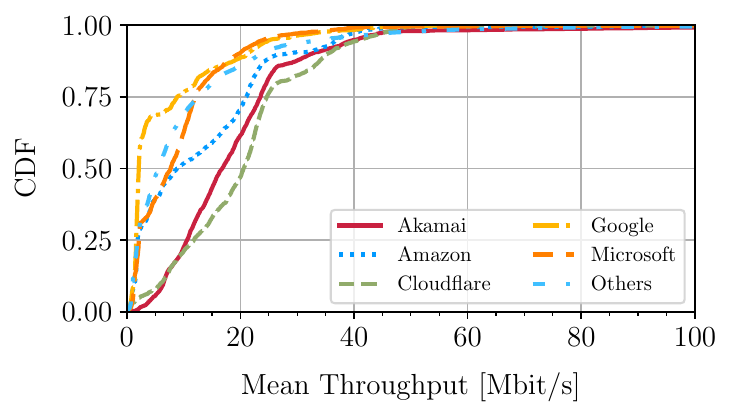}
    \caption{Vantage point in MUC.}
    \label{fig:pl_cdn_tp}
\end{subfigure}%
\begin{subfigure}{0.33\linewidth}
 \centering\includegraphics[width=\linewidth]{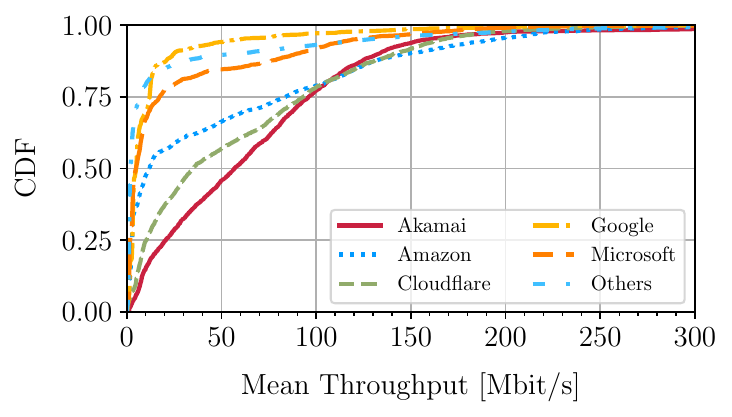}
    \caption{Vantage point in SFO.}
    \label{fig:dosf_cdn_tp}
\end{subfigure}%
\begin{subfigure}{0.33\linewidth}
 \centering\includegraphics[width=\linewidth]{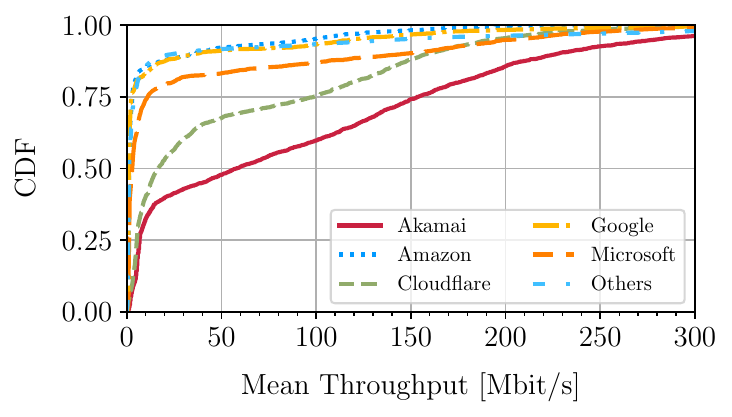}
    \caption{Vantage point in SGP.}
    \label{fig:dosg_cdn_tp}
\end{subfigure}%
\caption{Distribution of mean throughput measured for domains grouped by their CDN affiliation.}
\label{fig:cdn_impact}
\end{figure*}

\subsection{Impact of TCP Options}

To assess the impact of TCP option configuration on performance, we survey the CDF of mean throughput for each configuration, as shown in Figure\,\ref{fig:tcp_option_impact}.
For the VP in MUC, we observe two groups of distributions: \first{} configurations without enabled WS and \second{} configurations with enabled WS, while the latter indicates a significantly larger mean throughput. 
The same observation applies to measurements conducted with the VPs in SFO and SGP, which additionally show slightly increased mean throughput for downloads conducted with ECN, respectively SACK, compared to the baseline configuration. 

As the distribution of observed mean throughput does not show the explicit difference between two downloads of a measurement run, we calculate the share of measurements with an option configuration resulting in a specific speed-up compared to the baseline configuration.
Table\,\ref{tab:tcp_option_speedups} shows shares of downloads resulting in a specific speed-up gathered across all measurement iterations and vantage points.
As also observed for the CDFs of mean throughput, we find that measurements with TCP window scaling outperform measurements without window scaling. 
In particular,  window scaling results in a speed up larger than 1.5 for over \SI{60}{\percent} of conducted measurements.
Measurements with only ECN, respectively SACK enabled, mostly show comparable performance to the baseline measurements. 
Nearly \SI{70}{\percent} of measurements with ECN, respectively SACK, enabled result in a speed-up of +/- \SI{10}{\percent}. 

\subsection{TCP vs. QUIC}

To survey performance differences between TCP- and QUIC-based downloads, Table\,\ref{tab:tcp_option_speedups} shows the speed-ups by downloads conducted with \textit{aioquic} and \textit{quiche} compared to the throughput observed for TCP-BL and TCP-ALL.
Further, Table\,\ref{tab:tcp_option_speedups} compares the mean throughput observed for downloads conducted with \textit{quiche} to throughput observed for downloads conducted with \textit{aioquic}. 
We find that the mean throughput of downloads with \textit{quiche} exceeds the throughput of \SI{70}{\percent} of downloads with \textit{aioquic}, while mean throughput doubles for over \SI{35}{\percent} of downloads conducted with \textit{quiche}. 

Comparing the mean throughput of \textit{quiche} to TCP-BL downloads results in positive speed-ups for over \SI{80}{\percent} of measurements, while over \SI{65}{\percent} of measurements show a speed-up by more than \SI{50}{\percent}.
In comparison to TCP downloads with all options enabled, \textit{quiche} shows increased mean throughput for about \SI{70}{\percent} of samples, while the shares of larger speed-ups are significantly smaller, as observed for the comparison of \textit{quiche} and TCP-BL.

\subsection{CDN Impact}

To survey the impact of CDN hosting, we group measurement results according to the five considered giant CDNs and the sixth group, which includes all remaining domains.
Figure\,~\ref{fig:cdn_impact} shows the CDF of mean throughput per domain for considered CDNs.
The distribution of mean throughput shows that domains hosted by Cloudflare and Akamai provide the most significant shares of larger mean throughput. 
Domains hosted by Google and Microsoft show the least improved mean throughput compared to domains not hosted by one of the five giant CDNs. 

As mentioned in Section~\ref{sec:approach}, we conduct a warm-up download before all remaining option configurations to remove bias caused by potential caching of downloaded files. 
To further survey performance impacts by CDN hosting, we compare the distribution of mean throughput of warm-up runs to downloads conducted with the TCP baseline configuration, as shown in Figure\,\ref{fig:tcp_option_impact}.  
We observe that potential caching impacts are significantly larger for VPs in SFO and SGP, while such performance gains can be mainly traced back to targets hosted by Akamai, Amazon, and Cloudflare, which all provide edge caching services.
Such observation is confirmed by assessing throughput amplitudes between the warm-up run and the baseline download, as shown in Table\,\ref{tab:tcp_option_speedups}. 
While the majority of samples show comparable performance between the warm-up and the baseline, over \SI{15}{\percent} of warm-up samples result in a throughput decrease larger than \SI{30}{\percent}.  

In general, our measurements confirm the expectation that CDN hosting increases performance.
The degree of performance gain for each CDN varies between the three vantage points, which is reasonable since the vantage point location determines the nearest point of presence (PoP) of a CDN.

\section{Conclusion}

In this study, we conducted active Internet measurements with public web servers to assess the impact of TCP option usage, QUIC, and CDN hosting on connection performance. 

Our measurements show that TCP window scaling is crucial to increase throughput. 
Replacing TCP (using all options) with the QUIC implementation \textit{quiche} implies performance gain for over \SI{70}{\percent} of samples, while the used QUIC implementation also impacts measured performance. 
CDN hosting increases throughput for most considered CDNs compared to domains not hosted by one of the considered giant CDNs, while we observe varying performance depending on the vantage point.

For future work, we consider extending the introduced measurement pipeline to support additional protocol parameters, performance indicators, and analysis approaches like root cause analysis to determine throughput limitations.

\section*{Acknowledgments}

The authors would like to thank the anonymous reviewers for their valuable feedback. 
This work was partially funded by the German Federal Ministry of Education and Research under the project PRIMEnet (16KIS1370), 6G-life (16KISK002), and 6G-ANNA (16KISK107) as well as the German Research Foundation (HyperNIC, grant no. CA595/13-1). 
Additionally, we received funding by the Bavarian Ministry of Economic Affairs, Regional Development, and Energy as part of the project 6G Future Lab Bavaria and the European Union’s Horizon 2020 research and innovation program (grant agreement no. 101008468 and 101079774).%

\newpage
\bibliographystyle{IEEEtran}
\bibliography{lit}

\end{document}